\documentclass{article}
\usepackage{hiph-art}
\usepackage{graphicx}
\usepackage{amsmath}
\usepackage{amssymb}
\volnumber{00} \issuenumber{0} \edyear{0000}                             
\frompage{000} \topage{000}                                              
\recrevdate{7 November 2005}                                             
\def\rightmark{Multifragmentation and Symmetry Energy Studied with AMD}
\def\leftmark{A. Ono%
}
 
\newcommand{\calcium}[1][40]{{}^{#1}\mathrm{Ca}}

\title{Multifragmentation and Symmetry Energy Studied with
  Antisymmetrized Molecular Dynamics}
\authors{ 
{Akira Ono$^{1,2}$%
\index{Ono, A.} 
}\\[2.812mm]
{\normalsize
\hspace*{-8pt}$^1$ Department of Physics, Tohoku University, Sendai 980-8578, Japan\\[0.2ex] 
\hspace*{-8pt}$^2$ National Superconducting Cyclotron Laboratory,
Michigan State Univerisity, East Lansing, MI 48824, USA\\[0.2ex] 
}}
 
\abstract{ The antisymmetrized molecular dynamics (AMD) simulations
  suggest that the isospin composition of fragments produced
  dynamically in multifragmentation reactions is basically governed by
  the symmetry energy of low-density uniform nuclear matter rather
  than the symmetry energy for the ground-state finite nuclei.  After
  the statistical secondary decay of the excited fragments, the
  symmetry energy effect still remains in the fragment isospin
  composition, though the effect in the isoscaling parameter seems a
  very delicate problem.}


 
\makeindex
\begin{document}
 
\maketitle

\section{Introduction}\label{intro}
In medium energy heavy-ion collisions, many fragments are formed
almost simultaneously in an expanding and excited nuclear system,
which gives us an opportunity to investigate the nuclear matter
properties at various temperatures and densities, such as the nuclear
equation of state and the liquid-gas phase transition.  However,
fragments are formed in dynamically evolving system in most cases, and
therefore dynamical model calculations are necessary.

Antisymmetrized molecular dynamics (AMD) model
\cite{ONO-ppnp,ONOab,ONOh,ONOi,ONOj} respects several quantum features
in fragment formation reactions.  It uses a fully antisymmetrized
many-body wave function of Gaussian wave packets, which can describe
the ground state properties of nuclei reasonably well.  The
single-particle evolution in the mean field potential is described by
the the motion of the wave packet centroids and the stochastic quantum
branching process which respects the change of the shape of the
phase-space distribution, keeping the advantage of the molecular
dynamics that the fragments are formed with the wave packets localized
in phase space.  The two-nucleon collision effect is also treated as a
stochastic process.

The AMD simulations for heavy-ion collisions are useful not only to
explain the experimental data but also to know what kind of
information is reflected in the fragment formation.  In Refs.\
\cite{ONOk,ONOl}, we have analyzed the fragment yields in the AMD
simulations for multifragmentation reactions of the central collisions
of Ca isotopes at 35 MeV/nucleon, in order to see how the fragment
isospin composition is related to the symmetry energy term of the
effective interaction adopted in the calculation.  Isoscaling has been
observed in the fragment yield ratios from two reaction systems with
different proton-to-neutron ratios in the AMD results as well as in
the experimental data \cite{XU-isoscale} and in the predictions by
various statistical models \cite{TSANG-statmodel} and other dynamical
models \cite{SMF-isoscale,colonna-analytical,DORSO-isoscale}.  The
symmetry energy has turned out to be reflected in the fragment isospin
composition with almost no surface effect, which means that the
property of low-density uniform nuclear matter is responsible for the
fragment isospin composition rather than the symmetry energy for
finite nuclei.  The results for the primary fragments are discussed in
Sec.\ 2.

The fragments are recognized in the AMD simulations at a finite time
$t\sim300$ fm/$c$.  These primary fragments are excited with the
typical excitation energy of about 3 MeV/nucleon and they will decay
by emitting particles with a long time scale before they are detected
in experiments.  In order to compare the calculated results with the
experimental data, the secondary decay of primary fragments should be
considered by employing a statistical decay code.  The effect of the
secondary decay on the symmetry energy effects is studied in Sec.\ 3.
Some features observed in the primary fragments remain after the
secondary decay, which can be utilized to get the information of the
symmetry energy from the experimental data in principle.  However, the
symmetry energy effect in the isoscaling parameter is produced by a
very delicate cancellation of the effect in the width of the isotope
distribution and that in the mean isospin asymmetry of fragments.

\section{Symmetry Energy Effects in Primary Fragments}

\begin{figure}
\begin{center}
\includegraphics[width=0.4\textwidth]{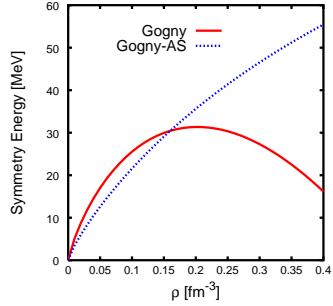}

\end{center}
\caption{\label{fig:symeng}
Density dependence of the symmetry energy of nuclear matter for
  the Gogny force and for the Gogny-AS force.}
\end{figure}

In this section, we discuss how the density-dependent symmetry energy
is reflected in the fragment isospin composition, by using the result
of AMD simulations of Refs.\ \cite{ONOk,ONOl} for
$\calcium[40]+\calcium[40]$, $\calcium[48]+\calcium[48]$, and
$\calcium[60]+\calcium[60]$.  We simulate collisions by boosting two
nuclei whose centers are separated by 9 fm and calculating the
dynamical evolution of each collision until $t=300$ fm/$c$.  The
primary fragments are recognized at $t=300$ fm/$c$ by the condition
that the two nucleons belong to the same fragment if the spatial
distance between them is less than 5 fm.  The calculations are done
with the Gogny force \cite{GOGNY} and the Gogny-AS force \cite{ONOk}
which are different in the density dependence of the symmetry energy
for nuclear matter as shown in Fig.\ \ref{fig:symeng}.

We have seen \cite{ONOk} that the fragment yield ratios
$Y_2(N,Z)/Y_1(N,Z)$ between different reaction systems satisfy
isoscaling
\begin{equation}
Y_{2}(N,Z)/Y_{1}(N,Z)\propto e^{\alpha N+\beta Z}.  \label{eq:isoscaling}
\end{equation}
Isoscaling is equivalent to the fact that the fragment yields of
different reaction systems $i$ are given by
\begin{equation}
Y_i(N,Z)=\exp[-K(N,Z)+\alpha_i N+\beta_i Z+\gamma_i],
\label{eq:YNZbyKNZ}
\end{equation}
where $\alpha_i$, $\beta_i$ and $\gamma_i$ are constants that depend
on the reaction system $i$, while $K(N,Z)$ is a function that is
independent of the reaction system.  By combining the fragment yields
from the three reaction systems, we can get the function $K(N,Z)$ for
a wide region of $(N,Z)$, even though the number of generated events
is not very large.

The obtained $K(N,Z)$ behaves very smoothly as a function of $N$ and
$Z$ \cite{ONOl}.  The shell and paring effects are weak in $K(N,Z)$
compared to the effects in the ground state binding energies.  We find
\cite{ONOl} that $K(N,Z)$ can be fitted very well by the functional
form
\begin{equation}
K(N,Z)=\xi(Z)N+\eta(Z)+\zeta(Z)\frac{(N-Z)^2}{N+Z}
\label{eq:KNZfit}
\end{equation}
for each $Z$.  The quantity $\zeta(Z)$ is related to the width of the
isotope distribution for the given $Z$, and it is expected to be
sensitive to the symmetry energy.

\begin{figure}
\begin{minipage}[t]{0.48\textwidth}
\includegraphics[width=\textwidth]{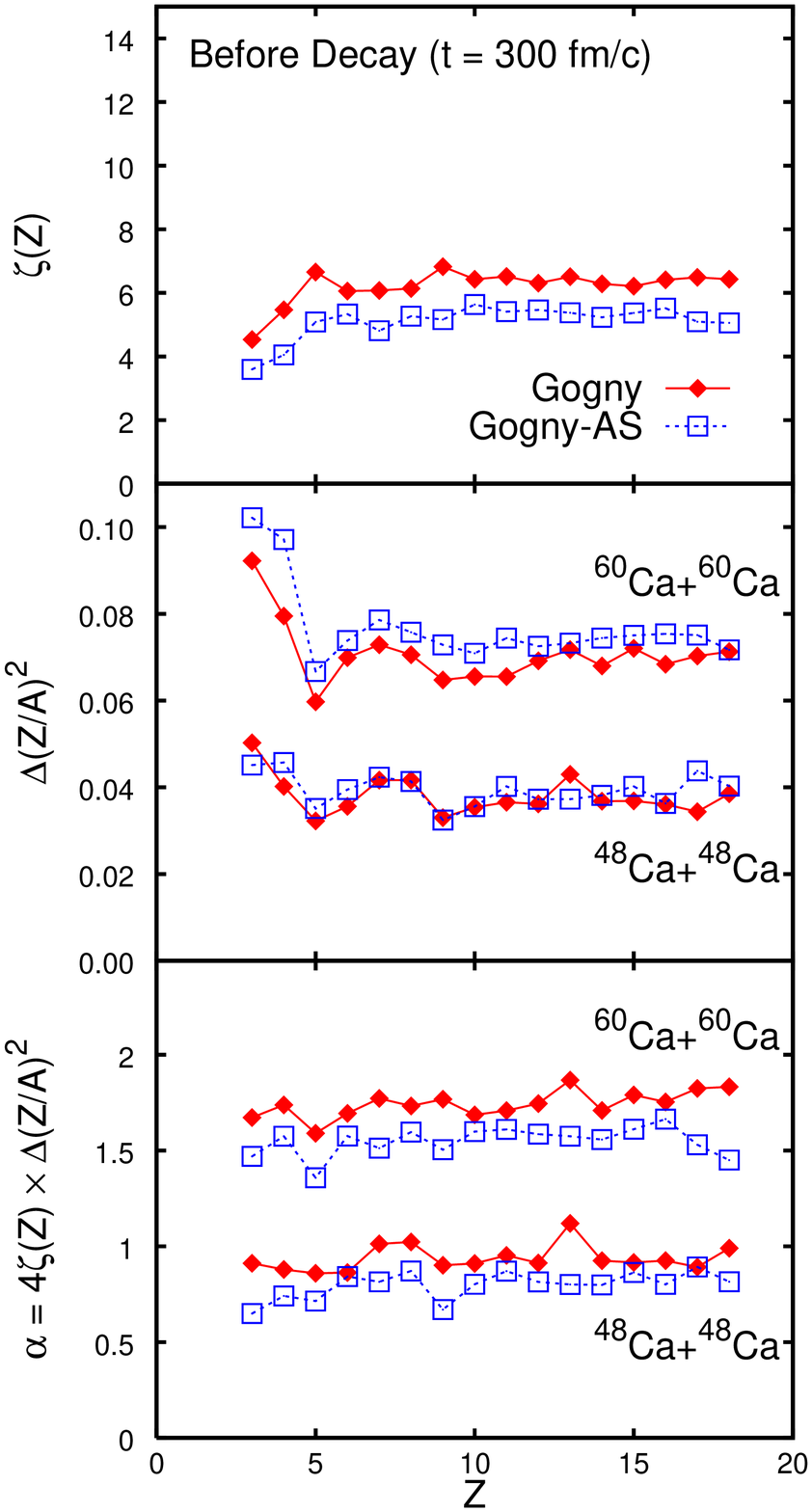}
\caption{\label{fig:zetanovz-dy} The three quantities $\zeta(Z)$
  (top), $\Delta(Z/A)^2$ (middle) and $\alpha$ (bottom) for the
  primary fragments.  The results with the Gogny and Gogny-AS forces
  are shown by filled diamonds and the open squares, respectively.}
\end{minipage}
\begin{minipage}[t]{0.48\textwidth}
\includegraphics[width=\textwidth]{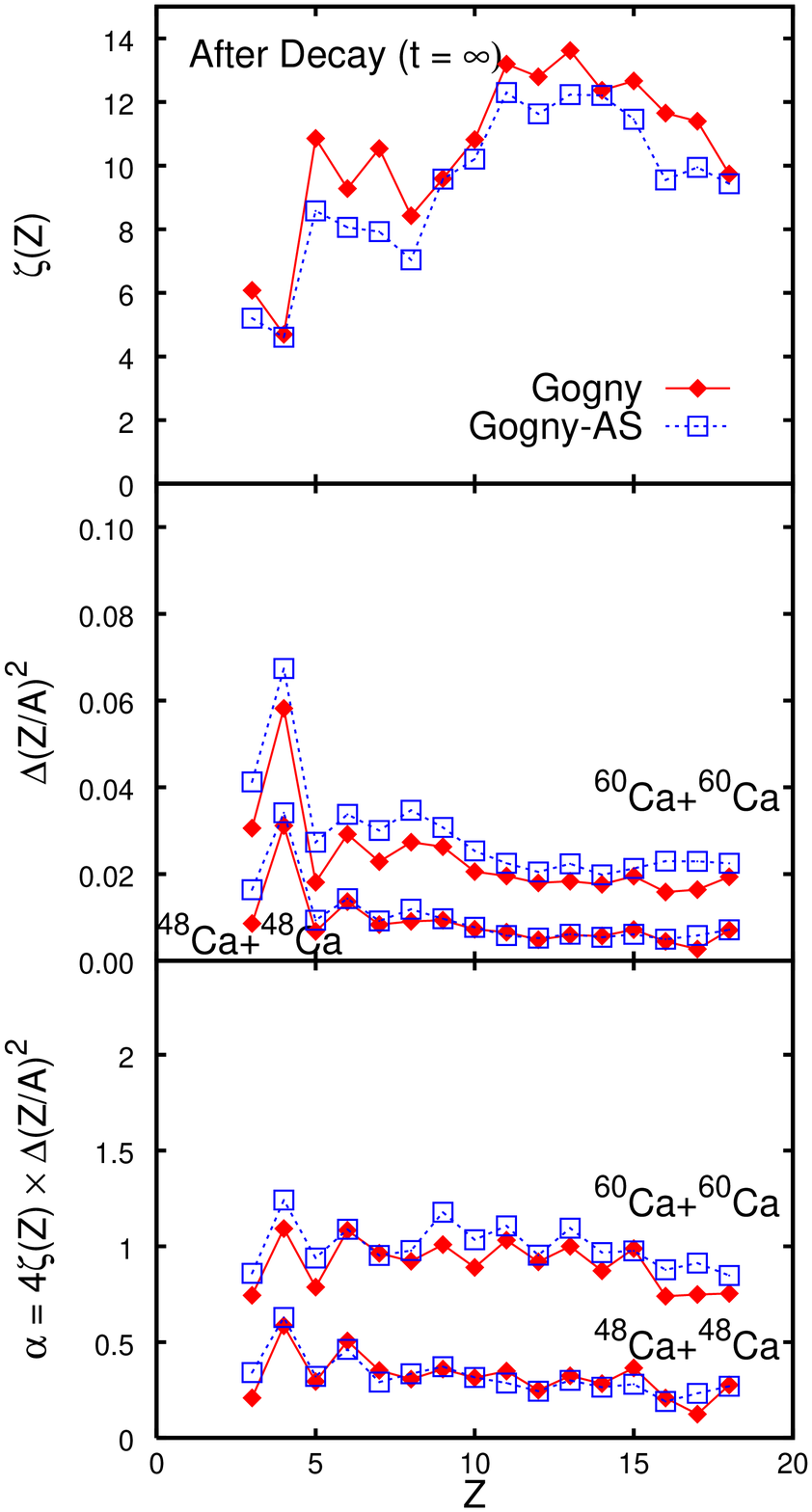}
\caption{\label{fig:zetanovz}
The same as Fig.\ \ref{fig:zetanovz} but for the final fragments.}
\end{minipage}
\end{figure}

The top panel of Fig.\ \ref{fig:zetanovz-dy} shows the obtained
parameter $\zeta(Z)$ as a function of the fragment proton number $Z$
for the Gogny force and the Gogny-AS force.  We find that the
$Z$-dependence of $\zeta(Z)$ is very weak for $Z\gtrsim5$, which is
not consistent with the idea of the usual equilibrium of primary
fragments.  In fact, if the fragments were in thermal and chemical
equilibrium, $\zeta(Z)$ would be related to the symmetry energy for
the fragment nuclei as
$\zeta(Z)=(c_{\text{v}}+c_{\text{s}}A^{-1/3})/T$, where the volume and
surface symmetry energy coefficients satisfy
$c_{\text{v}}\approx-c_{\text{s}}$ for the ground state nuclei.  This
size dependence of the symmetry energy is not compatible with the very
weak $Z$-dependence of $\zeta(Z)$ for $Z\gtrsim5$ \cite{ONOl}.

It may be rather surprising that we can find almost no surface effect
even though we are looking at relatively small fragments.  There can
be several possible explanations \cite{ONOl}.  One of them is
associated with the fact that fragments are not isolated when they are
formed.  When the density fluctuation is developing from a uniform low
density matter, the fragments are still interacting with attractive
force through their surfaces.  Therefore, the surface free energies
can be expected to be smaller for these fragments than for the
isolated fragments.  Independent of the physical origin for the
weakening of the surface symmetry free energy, it suggests that the
volume quantity, which is the same as that in the infinite nuclear
matter, can be directly obtained by the analysis of the fragmentation
results even though the produced fragments are not very large.

If we adopt the interpretation that $\zeta$ is related to the symmetry
energy $C_{\text{sym}}(\rho)$ for the uniform matter at a certain
density $\rho$ by
\begin{equation}
\zeta=\frac{C_{\text{sym}}(\rho)}{T},
\label{eq:zetaCsymT}
\end{equation}
we can utilize the symmetry energy effect observed in the top panel of
Fig.\ \ref{fig:zetanovz-dy} to derive $\rho$ and $T$ \cite{ONOk}.  In
order to explain the ratio of $\zeta$ for the Gogny force and the
Gogny-AS force which are different in the density dependence of the
symmetry energy as shown in Fig.\ \ref{fig:symeng}, the density should
be $\rho\sim\frac{1}{2}\rho_0$.  Furthermore, from the absolute values
of $\zeta$ and $C_{\text{sym}}(\frac{1}{2}\rho_0)$, the temperature
should be $T\sim3.4$ MeV.  This condition of the density and the
temperature is reasonable as the condition for the fragment formation,
which is necessary for the justification of Eq.\ \eqref{eq:zetaCsymT}.

If the fragment yields from different reaction systems satisfy
isoscaling, the isoscaling parameter $\alpha$ is related to $\zeta(Z)$
by \cite{ONOk,ONOl}
\begin{equation}
  \alpha=4\zeta(Z)\times\Delta(Z/A)^2,
\label{eq:linearrel}
\end{equation}
where
\begin{equation}
  \Delta(Z/A)^2=(Z/\bar{A}_1(Z))^2-(Z/\bar{A}_2(Z))^2
\end{equation}
represents the difference of the mean isospin asymmetry of fragments
for each given $Z$ between the two reaction systems.  We will choose
the $\calcium[40]+\calcium[40]$ system as the system 1 in this paper.
Equation \eqref{eq:linearrel} can be derived from Eq.\
\eqref{eq:YNZbyKNZ} by employing Eq.\ \eqref{eq:KNZfit}, and thus it
is not relevant how $\zeta(Z)$ is related to the symmetry energy.  It
should be noted that $\alpha$ and $\Delta(Z/A)^2$ depend on the
reaction systems while $\zeta(Z)$ does not, and that $\zeta(Z)$ and
$\Delta(Z/A)^2$ are functions of $Z$ while $\alpha$ should be
independent of $Z$.

The middle panel of Fig.\ \ref{fig:zetanovz-dy} shows $\Delta(Z/A)^2$
as a function of $Z$ when the $\calcium[60]+\calcium[60]$ (and
$\calcium[48]+\calcium[48]$) system is chosen as system 2.  The
results with the two different symmetry energy terms are shown.  The
symmetry energy effect is clearly seen in $\Delta(Z/A)^2$ as well as
in $\zeta(Z)$, though the effect is in the opposite direction.  The
bottom panel shows $4\zeta(Z)\times\Delta(Z/A)^2$ which is the product
of the top and middle panels and should coincide with the isoscaling
parameter $\alpha$.  The product is actually almost independent of
$Z$, which is consistent with Eq.\ \eqref{eq:linearrel}.  The symmetry
energy effect in $\alpha$ is similar to the effect in $\zeta(Z)$, but
it has been produced by the cancellation of the effect in $\zeta(Z)$
and the effect in $\Delta(Z/A)^2$, with the former being larger than
the latter.

\section{Symmetry Energy Effects in Final Fragments}

The statistical decay of primary fragments is calculated by using the
code \cite{MARUb} based on the sequential binary decay model by
P\"uhlhofer \cite{PUHLHOFER}.  The code employed in the present work
also takes account of the emission of composite particles not only in
their ground states but also in their excited states with the
excitation energy $E^*\le 40$ MeV.  The experimental information is
incorporated for known levels of $A\lesssim28$ nuclei, while the
Fermi-gas level density is assumed otherwise with the level density
parameter $a=A/(8\ \text{MeV})$.

For the final fragments after the statistical decay, the quantities
$\zeta(Z)$ and $\Delta(Z/A)^2$ can be defined in the same way as for
the primary fragments.  Equation \eqref{eq:linearrel} holds for the
final fragments as well if isoscaling is good and Eq.\
(\ref{eq:KNZfit}) is approximately valid for the final fragments.
However, $\zeta(Z)$ for the final fragments is not necessarily related
to the symmetry energy directly in a simple way.

The top and middle panels of Fig.\ \ref{fig:zetanovz} show $\zeta(Z)$
and $\Delta(Z/A)^2$, respectively, calculated for the final fragments.
The absolute values of $\zeta(Z)$ and $\Delta(Z/A)^2$ change from
those in Fig.\ \ref{fig:zetanovz-dy} for the primary fragments.
Nevertheless, the effect of the symmetry energy term in $\zeta(Z)$ and
$\Delta(Z/A)^2$ is similar to what we have observed for the primary
fragments.  This means that the symmetry energy effect in the
dynamical stage of the reaction is observable in principle even after
the secondary decay of fragments.

However, in the isoscaling parameter $\alpha$ which is shown as
$4\zeta(Z)\times\Delta(Z/A)^2$ in the bottom panel of Fig.\
\ref{fig:zetanovz}, the symmetry energy effect is not clearly seen.
This is because the symmetry energy effect in $\zeta(Z)$ is largely
canceled by that in $\Delta(Z/A)^2$ to give this small effect in
$\alpha$.  We expect that the situation of such cancellation may be
sensitive to the details of the statistical decay code, and therefore
it is not easy to predict the symmetry energy effect in $\alpha$ for
the final fragments at the current stage.

\section{Summary}

The isospin composition of fragments produced by the AMD simulations
is analyzed to see how the symmetry energy is reflected.  The studied
quantities $\zeta(Z)$, $\Delta(Z/A)^2$ and $\alpha$ for the primary
fragments are sensitive to the symmetry energy term in the effective
interaction, and the calculated results suggests that the symmetry
energy of uniform nuclear matter at a reduced density
$\rho\sim\frac{1}{2}\rho_0$ is directly responsible for the isospin
composition of the primary fragments even though the fragments are of
finite size.  A statistical decay calculation shows that the symmetry
energy effects in $\zeta(Z)$ and $\Delta(Z/A)^2$ remain after the
secondary decay of the primary fragments.  However, the symmetry
energy effect in $\alpha$ for the final fragments requires a careful
study because it is given by a delicate cancellation of the effect in
$\zeta(Z)$ and the effect in $\Delta(Z/A)^2$.

\section*{Acknowledgments}
The main part of this talk was based on the collaboration with
M. B. Tsang, W. G. Lynch, P. Danielewicz and W. A. Friedman, which was
supported by Japan Society for the Promotions of Science and the US
National Science Foundation under the U.S.-Japan Cooperative Science
Program (INT-0124186), by the High Energy Accelerator Research
Organization (KEK) as the Supercomputer Project, and by grants from
the US National Science Foundation, PHY-0245009, PHY-0070161 and
PHY-01-10253, and a Grant-in-Aid for Scientific Research from the
Japan Ministry of Education, Science and Culture.  The work was also
partially supported by RIKEN as a nuclear theory project.

\vfill\eject
\end{document}